\documentclass{WileyMSP-template}

\begin{document}

\pagestyle{fancy}
\rhead{ }

\title{Toward Functionalized Ultrathin Oxide Films:\\ the Impact of Surface Apical Oxygen}
This is the pre-peer reviewed version of the following article: Adv. Electron. Mater. 2021, 2101006, which has been published Open Access in final form  at https://doi.org/10.1002/aelm.202101006. \\
\maketitle

\newcommand{\etal}{\textit{et~al.} }

\newcommand{\fref}[1]{Figure \ref{#1}}

\author{Judith Gabel}
\author{Matthias Pickem}
\author{Philipp Scheiderer}
\author{Lenart Dudy}
\author{Berengar Leikert}
\author{Marius  Fuchs}
\author{Martin St\"ubinger}
\author{Matthias Schmitt}
\author{Julia K\"uspert}
\author{Giorgio Sangiovanni}
\author{Jan M.~Tomczak$^*$}
\author{Karsten Held}
\author{Tien-Lin Lee$^*$}
\author{Ralph Claessen}
\author{Michael Sing$^*$}

\begin{affiliations}
Judith Gabel, Tien-Lin Lee\\
Diamond Light Source, Didcot, OX11 0DE, United Kingdom\\
Email Address: tien-lin.lee@diamond.ac.uk

Matthias Pickem, Jan M.~Tomczak, Karsten Held\\
Institute of Solid State Physics, TU Wien,  Vienna, Austria\\
Email Address: jan.tomczak@tuwien.ac.at

Philipp Scheiderer, Berengar Leikert, Martin St\"ubinger, Matthias Schmitt, Julia K\"uspert, Ralph Claessen, Michael Sing\\
Physikalisches Institut and W\"urzburg-Dresden Cluster of Excellence ct.qmat, Universit\"at W\"urzburg, 97074 W\"urzburg, Germany\\
Email Address: sing@physik.uni-wuerzburg.de

Lenart Dudy\\
Synchrotron SOLEIL, 91190 Saint-Aubin, France

Marius  Fuchs, Giorgio Sangiovanni\\
Institut für Theoretische Physik and W\"urzburg-Dresden Cluster of Excellence ct.qmat,  Universit\"at W\"urzburg, 97074 W\"urzburg, Germany

\end{affiliations}



\begin{abstract}
Thin films of transition metal oxides open up a gateway to nanoscale electronic devices beyond silicon  characterized by novel electronic functionalities. 
While such  films are commonly prepared in an oxygen atmosphere, they are typically considered to be ideally terminated with the stoichiometric composition. Using the prototypical correlated metal SrVO$_3$ as an example, it is demonstrated that this idealized description overlooks an essential ingredient: oxygen adsorbing at the surface apical  sites. 
The oxygen adatoms, which persist  even in an ultrahigh vacuum environment, are shown to  severely affect the intrinsic electronic structure of a transition metal oxide film. 
Their presence leads to the formation of an electronically dead surface layer but also alters the band filling and the electron correlations in the thin films.
These findings highlight that it is important to take into account surface apical oxygen or  -- mutatis mutandis-- the specific oxygen configuration imposed by a capping layer to predict the behavior of ultrathin films of transition metal oxides near the single unit-cell limit.

\end{abstract}

Their extraordinary electronic properties have made transition metal oxides (TMOs) promising candidates for next-generation  electronic device and memory structures \cite{ahnElectricFieldEffect2003, heberMaterialsScienceEnter2009}. 
One prominent example is the metal-insulator transition in correlated oxides, which may be exploited to form a new class of electronic devices – referred to as Mottronics \cite{sonHeterojunctionModulationdopedMott2011, scheiderer_tailoring_2018}. A Mott transistor, e.g., employs the electronic phase transition between the states of a  Mott insulator and a correlated metal as an on-off switch,  which can be far more effective and robust than merely manipulating the carrier density as is done in conventional semiconductor devices \cite{Newns1998,guMetalinsulatorTransitionInduced2013, scheiderer_tailoring_2018}.\\
To realize these novel devices and, in particular, to miniaturize them, TMOs need to be prepared in the form of thin films, which, among many other advantages, allows unstable phases of bulk materials to be synthesized and strain to be tuned.
In addition, the film thickness itself can be  exploited as a tuning parameter in navigating the electronic properties of TMOs through their phase diagrams.
For instance, a Mott metal-insulator transition can be triggered in many correlated oxides when the two-dimensional limit is approached.
\cite{scherwitzlMetalInsulatorTransitionUltrathin2011, toyotaThicknessdependentElectronicStructure, xiaCriticalThicknessItinerant2009, rondinelliElectronicPropertiesBulk2008, schutzDimensionalityDrivenMetalInsulatorTransition2017, guMetalinsulatorTransitionInduced2013,yoshimatsu_dimensional_2010} \\
While these ultrathin films are very sensitive at any modification to their surfaces, they are typically described as slabs with the ideal bulk stoichiometry throughout  and an abrupt surface\cite{yoshimatsu_metallic_2011, scherwitzlMetalInsulatorTransitionUltrathin2011,schutzDimensionalityDrivenMetalInsulatorTransition2017}, leading to surface cations with a lower coordination. However, if the transition metal ion has a tendency to oxidize further from its bulk oxidation state, the truncated surface may host adsorption sites that attract additional oxygen to restore the bulk ligand coordination for the transition metal ions in the surface layer. 
Indeed, oxygen adsorption has been observed on various TMO-surfaces such as perovskite SrVO$_3$ \cite{takizawa_coherent_2009,yoshimatsu_metallic_2011,okada_quasiparticle_2017}, La$_{5/8}$Ca$_{3/8}$MnO$_3$ \cite{ fuchigamiTunableMetallicityCa2009}, SrIrO$_3$ \cite{schutzElectronicStructureEpitaxial2020},  LaTiO$_3$ \cite{scheiderer_tailoring_2018} and LaVO$_3$ (see Section 6) surfaces,   rutile  VO$_2$ \cite{wahilaBreakdownMottPhysics2020} and corundum V$_2$O$_3$ \cite{Window2011,Feiten2015,Lantz2015,wahilaBreakdownMottPhysics2020} surfaces. 
However, the impact of the surface apical oxygen on the electronic structures of the thin films has been rarely discussed in the previous publications.\\
To investigate this impact, we use the prototypical correlated metal SrVO$_3$ (SVO),\cite{fujimori_evolution_1992,inoue_systematic_1995,sekiyama_mutual_2004, sekiyama_mutual_2004,rozenberg_low_1996, pavarini_how_2005,lee_dynamical_2012,tomczak_gwdmft_2012,taranto_comparing_2013,tomczak_asymmetry_2014,roekeghem_screened_2014} a material which is attractive as highly conductive bottom electrode and transparent conductor \cite{moyer_highly_2013,zhang_correlated_2016,Si_trans_cond_2020} as well as a promising channel material for the realization of a Mott transistor \cite{zhong_electronics_2015}. 
By means of in situ photoelectron spectroscopy combined with realistic electronic structure calculations
we demonstrate that  the overoxidized surface, which forms inevitably under the growth conditions and persists in ultrahigh vacuum,  severely affects the electronic and atomic structures of the ultrathin TMO film.
We show that the apical oxygen not only leads to the formation of regions of an electronically and magnetically dead layer at the surface but also alters the band filling and the electronic correlations of the thin films, rendering the conventional consideration of  a simple stoichiometric slab invalid. 
In more general terms, our results showcase  the importance of the oxygen configuration at the surface of transition metal oxide thin film. 
With the oxygen configuration being not only affected by apical oxygen but also by potential capping layers, our findings are highly relevant to the design and operation of integrated TMO-based devices with an active layer thickness approaching the unit cell regime, such as the emerging Mott transistors.

\section{Surface overoxidation of thick SrVO$_3$ films}
\begin{figure*}[hbpt]
\includegraphics[width = \linewidth]{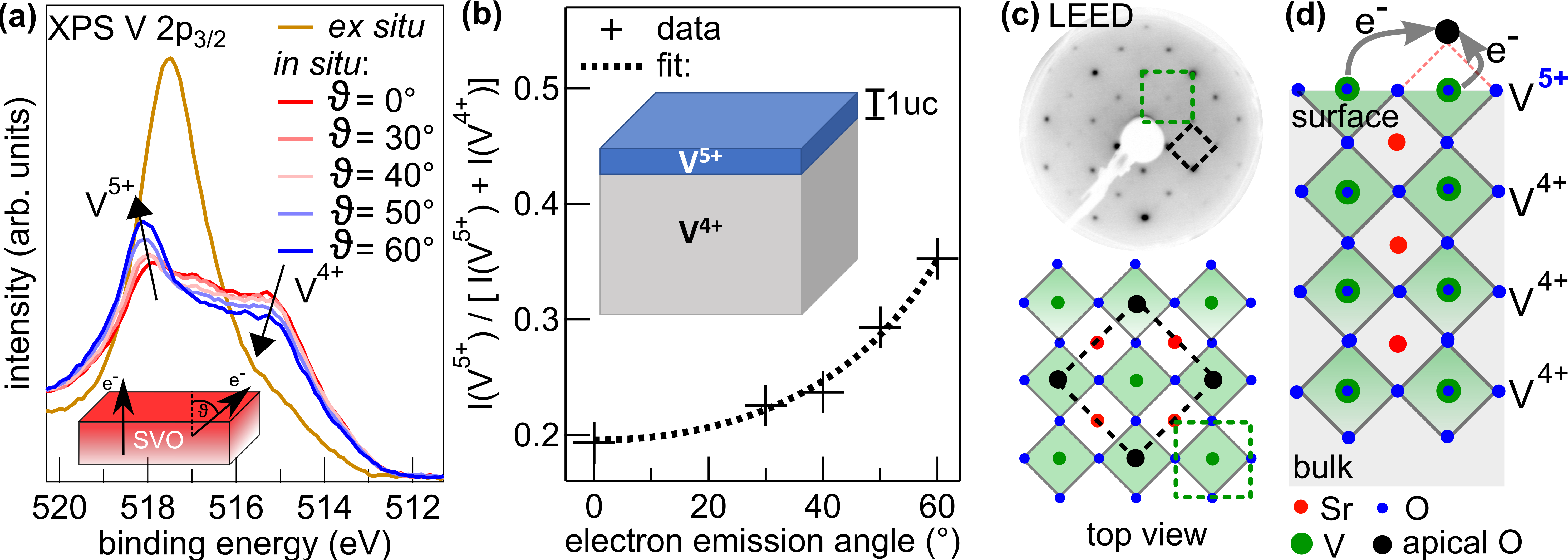}
\caption{
\label{FigXPSthickfilm}
a) V\,2$p_{3/2}$ spectra of  a 75\,uc thick SVO film exposed to air (\emph{ex situ}) are dominated by  a strong V$^{5+}$ signal evidencing the oxidation of the SVO film.  
XPS on \emph{in vacuo} (\emph{in situ}) handled SVO films at different electron emission angles $\vartheta$ detects a V$^{5+}$ signal that emanates from the surface, even without exposure to air.  
b) Model fit of the angle-dependent relative V$^{5+}$ content $I$(V$^{5+}$)/[$I$(V$^{4+}$)+$I$(V$^{5+}$)]. The best match is achieved with a 
 $d^1$ occupancy throughout the film  except for a single dead layer at the very surface with $d^0$ occupancy.
c) Top: The LEED pattern of a thick SVO film exhibits a $\sqrt{2}\times\sqrt{2}$ R(45$^{\circ}$) surface reconstruction consistent with the ordered adsorption of  oxygen ions at apical sites as detected in an STM study\cite{okada_quasiparticle_2017}.
Bottom: Structural model of the SVO film surface decorated with apical oxygen. 
d) Structural model of the SVO film showing the additional oxygen ions draining electrons from the film surface.   
}
\end{figure*}

The nominal $3d^1$ occupancy of V in SrV$^{4+}$O$_3$ poses a challenge for the sample preparation, because the most stable vanadium valence, which corresponds to a noble gas configuration, is not V$^{4+}$ but V$^{5+}$ making SVO films prone to overoxidation.
This becomes apparent in Figure \ref{FigXPSthickfilm}a, where  x-ray photoelectron spectroscopy (XPS) was employed to record the  V\,$2p_{3/2}$ core level spectra of an epitaxial 75\,uc thick SVO film  grown by pulsed laser deposition (PLD) with a nominal VO$_2$ surface termination.
For the sample exposed to air prior to the XPS measurement the spectrum (labelled \emph{ex situ})  is dominated by a strong peak around a binding energy of   518\,eV that is assigned to V$^{5+}$ ions by comparison to reference data of different vanadium oxide compounds \cite{silversmit_determination_2004}.
However, even for films analyzed  immediately after the PLD growth without an exposure to air, pentavalent vanadium is detected \cite{takizawa_coherent_2009}, as is seen from the \emph{in situ} spectra in
Figure \ref{FigXPSthickfilm}a. 
The spectra   were recorded at different electron emission angles $\vartheta$ to check whether this V$^{5+}$ signal originates from the SVO film surface.  The V$^{5+}$ component is stronger at higher $\vartheta$, \emph{i.e.}, in more surface sensitive measurement geometries, which  suggests that the over-oxidation occurs near the SVO surface. 
For a quantitative account, the angle-dependence of the  relative V$^{5+}$ content \\ $I$(V$^{5+}$)/[$I$(V$^{4+}$)+$I$(V$^{5+}$)] was fitted (see Figure \ref{FigXPSthickfilm}b) within a microscopic model taking into account the relative V$^{5+}$ content in each VO$_2$ layer of the film and the depth-dependent damping of the photoelectrons  (see Section 2 in Supporting Information for details).  We found that the V$^{5+}$ component stems almost exclusively from the topmost layer  while the subsurface layers only contribute to the V$^{4+}$ component.  \\
Complementary structural information is provided by low-energy electron diffraction (LEED). The \emph{in situ} observed diffraction pattern of a thick film  in Figure~\ref{FigXPSthickfilm}c exhibits a $\sqrt{2}\times\sqrt{2}$ R(45$^{\circ}$) reconstruction. The associated reciprocal and direct space surface unit cells  are marked in the pattern and in the structural model of the surface  by dashed green and black squares for the cubic perovskite and the reconstruction, respectively. The same surface reconstruction was reported in the literature for similarly grown SVO films\cite{takizawa_coherent_2009,yoshimatsu_metallic_2011,okada_quasiparticle_2017}, and a recent scanning tunneling microscopy (STM) study by Okada \etal \cite{okada_quasiparticle_2017} identifies adsorbed oxygen ions as its microscopic origin. They find that additional oxygen ions occupy half of the apical sites above the surface vanadium ions, thereby closing every other VO$_6$ octahedron.
The oxygen presumably adsorbs during the growth of the samples which is performed in an oxygen background atmosphere. This microscopic picture is consistent with the over-oxidation of the topmost VO$_2$ layer inferred from the modelling in Figure \ref{FigXPSthickfilm}b since every excess oxygen ion drains two electrons from the V\,3$d$ band and thereby (nominally) generates two V$^{5+}$ ions, as sketched in the structural model in  Figure~\ref{FigXPSthickfilm}d \cite{okada_quasiparticle_2017}.
Therefore, we conclude that a complete electronically dead ($d^0$) surface layer forms on a thick SVO film terminated with an apical oxygen induced superstructure while the rest of the film  remains virtually unaffected.

\section{Evolution of apical oxygen coverage with film thickness}
\begin{figure*}[hbpt]
\includegraphics[width = \linewidth]{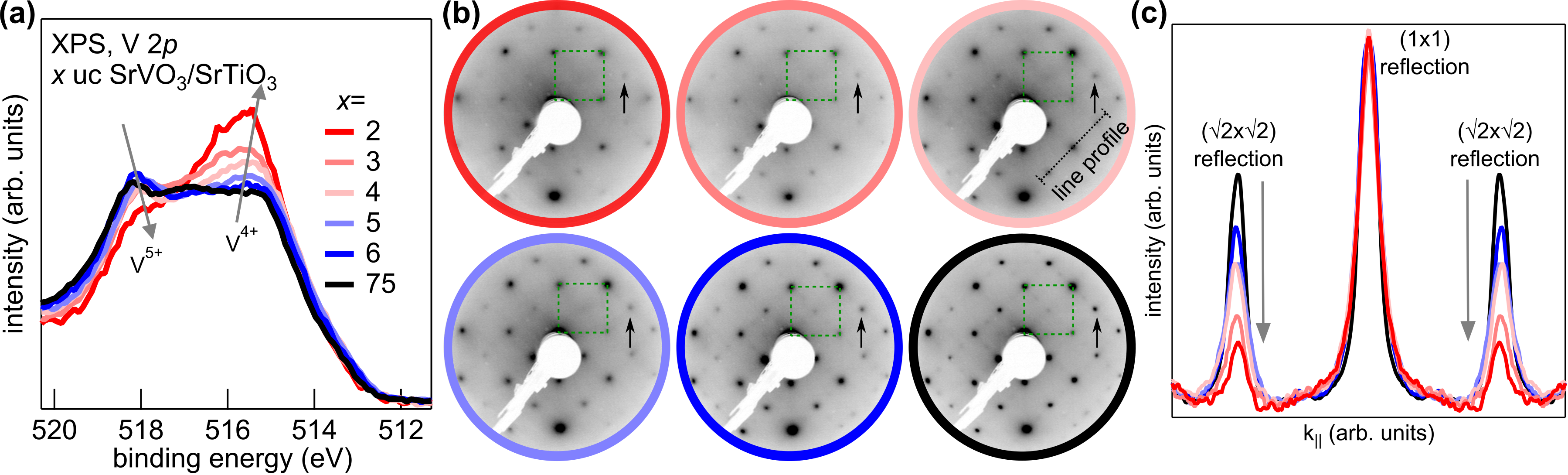}
\caption{
\label{FigThicknessDep}
a) V\,2$p_{3/2}$ spectra of SVO films with thicknesses across the dimensional crossover show the V$^{5+}$ signal  decreasing for thinner films. b) Corresponding LEED patterns where the diffraction pattern of the cubic perovskite unit cell is marked by dashed squares. The $\sqrt{2}\times\sqrt{2}$ R(45$^\circ$)  surface reconstruction (see spot highlighted by arrows) is present on every sample.   
c) Intensity of LEED spots along the line profile indicated in b). The line profiles are symmetrized with respect to the (1$\times$1) reflection. At smaller SVO thicknesses  the reflections assigned to the surface reconstruction lose intensity with respect to the diffraction pattern of the cubic perovskite.
  }
\end{figure*}

With the surface overoxidation essentially only affecting  the valence
of the topmost SVO layer, it can  be mostly neglected for the physical properties of thick SVO film. 
When the SVO film thickness is reduced to only a few unit cells, the adsorbed oxygen ions may, however, have a strong bearing  on the electronic properties of the films.  
In the next step, we thus investigate  ultrathin SVO films of 2 to 6\,uc  thicknesses \emph{in situ} immediately after their growth. 
Figure~\ref{FigThicknessDep}a depicts the corresponding XPS V\,2$p_{3/2}$ spectra together with that of the 75\,uc thick film, normalized  to the integral V\,2$p$ spectral weight. The spectra exhibit a systematic trend:
The intensity of the V$^{5+}$ component increases with the film thickness and saturates above 6\,uc SVO. 
To back up the observation in the V\,2$p$ spectra, we present the corresponding LEED patterns in Figure \ref{FigThicknessDep}b, where the $\sqrt{2}\times\sqrt{2}$ R(45$^{\circ}$) reconstruction ascribed to apical oxygen adsorption is present on all samples. 
However, compared to the $1\times1$ spots, the fractional order reflections (marked by the arrows) lose their intensity with decreasing film thickness. 
This becomes apparent in Figure \ref{FigThicknessDep}c in which the intensity of the LEED pattern is analyzed along the dotted line in \ref{FigThicknessDep}b. 
With the line profile normalized to the peak intensity of the  $1\times1$  reflection, the  $\sqrt{2}\times\sqrt{2}$ R(45$^{\circ}$) spots clearly weaken  with decreasing SVO thickness, indicating a declining surface coverage with adsorbed oxygen, consistent with the trend revealed by XPS for the V$^{5+}$ component in Figure \ref{FigXPSthickfilm}a. 
Note that the intensity changes  in the diffraction pattern do not permit 
a quantitative estimate of the total amount of adsorbed oxygen, as  LEED detects only the ordered apical atoms incorporated in the superstructure and is not sensitive to   oxygen adatoms randomly occupying the remaining apical sites. 
Nonetheless, since the apparent trend detected in the LEED patterns mirrors that observed in the V\,2$p_{3/2}$ spectra, we arrive at the conclusion that  the coverage of adsorbed apical oxygen and hence the fraction of    V$^{5+}$ ions  gradually reduce with decreasing SVO film thickness.\\

\section{Effect of the surface overoxidation  -- theory}

\begin{figure}[hbt]
\includegraphics[width=\linewidth]{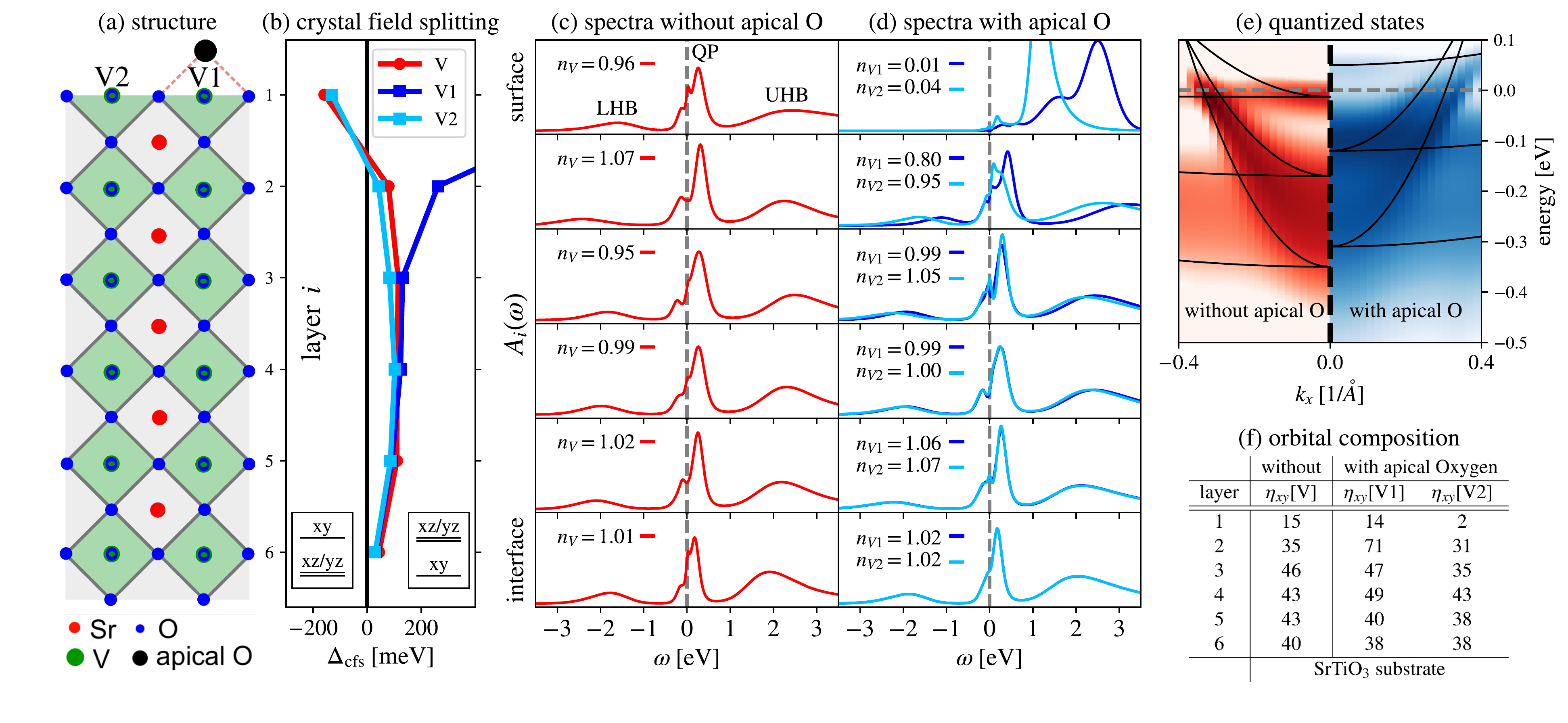}
\centering
\caption{
\label{Figtheory}
a) Sketch of the calculated 6\,uc heterostructure.
b) Crystal field splitting  $\Delta_{\mathrm{cfs}}$ of the vanadium t$_{2g}$ orbitals from DFT.
 With an apical oxygen (black dot) there are two inequivalent vanadium sites ``V1'' and ``V2'' per layer, without they are equivalent and denoted by ``V''. The corresponding local orbital energy levels are illustrated in the two insets.
 The crystal field splitting of ``V1'' in the two topmost layers is a direct result of the apical oxygen. At the surface layer the crystal field splitting for ``V1'' is 1\,eV  which is outside the drawn energy range. 
c,d) Layer($i$)-resolved DMFT spectral functions $A_i(\omega)$ for 6\,uc SVO (c) without 
and (d) with apical oxygen.
e) Quantized states along the $k_x$ direction ($k_y=0$) in the structures with and without apical oxygen (lines are obtained from parabolic fits to the second derivative of the shown intensity map).
f) Orbital composition of the different vanadium sites as the percentage of the $xy$ occupation  $\eta_{xy}$=$\frac{n_{xy}}{n_{xy} + n_{xz} + n_{yz}}$ for the layers from top to bottom. The proportion of the $xz/yz$ orbital can be obtained \emph{via} $\eta_{xz/yz} = {(1-\eta_{xy})}/{2}$.
}
\end{figure}

With the  adsorbed oxygen  draining electrons from the SVO films, we can expect their intrinsic low-energy electronic properties  to be strongly modified with respect to fictitious bulk-like films with an ideal abrupt surface. 
To address this question, we perform 
realistic electronic structure calculations for 6\,uc SVO on a STO substrate, using density functional theory (DFT) plus dynamical mean field theory (DMFT) with DFT structural relaxation.
For computational details see the Methods section, and Section 5 in the Supporting Information for
films of different thicknesses.
Without apical oxygen, the surface layer shows a DFT crystal-field splitting of the V\,3$d$ shell between the (in-plane)  $xy$-orbital and the (degenerate) $xz/yz$ orbitals of $\Delta_{\rm cfs}=E_{xz/yz}-E_{xy}=-155$meV, see curve ''V'' in \fref{Figtheory}b. 
This splitting results  
from two competing effects: (i) the topmost 
vanadium atoms lack an apical oxygen to form a complete octahedron, which for an ideal cut-off octahedron would yield a value of
$\Delta_{cfs}=-370$meV;

(ii) the latter is slightly compensated 
by the structural relaxation as well as by the tensile strain imposed by the STO substrate which enhances the distance of the $xy$ orbital lobes to the 
in-plane oxygens. \cite{pickemZoologySpinOrbital}
Accordingly, for the subsurface layers, where all oxygen octahedra are complete, $\Delta_{\rm cfs}$ is positive as seen in \fref{Figtheory}b.
Towards the STO substrate this effect is largely compensated by the (out-of-plane) lattice mismatch between SVO and STO.

Next, many-body effects are included using DMFT. Without apical oxygen, the spectral functions in Figure~\ref{Figtheory}c are metallic for all layers, with the three-peak structure, consisting of lower Hubbard band, quasiparticle peak (with some fine structure) and upper Hubbard band, being prototypical for a correlated metal such as (bulk) SVO. All vanadium sites
can be described with a nominal $d^1$ configuration (\emph{i.e.}, V$^{4+}$). However, because of the layer-dependent crystal field splitting $\Delta_{\rm cfs}$  the orbital compositions differ vastly, see \fref{Figtheory}f.

We now turn to the system with apical oxygen, where we consider the idealized 
experimental $\sqrt{2}\times\sqrt{2}$ cell with  50\% apical oxygen coverage. 
There are two inequivalent vanadium sites per layer, with and without apical oxygen,  denoted ``V1'' and ``V2'', respectively (see  \fref{Figtheory}a).
The crystal field splitting for ``V2'' is similar to the case without oxygen in \fref{Figtheory}b. The splitting for vanadium ions ``V1'', however, is dramatically changed, even inverted, near the surface. 

In all, apical oxygen-induced changes in the crystal and electronic structure are confined to the uppermost two layers, see the  DMFT spectral functions in Figure~\ref{Figtheory}d. Whereas the third to sixth layer  are practically identical to the case without apical oxygen in Figure~\ref{Figtheory}c, the  top layer shows dramatic qualitative changes: it is insulating and contains only pentavalent vanadium with a $d$ occupation $n_{V_{1,2}}\approx 0$, \emph{i.e.}, no occupied states below the Fermi level at 0\,eV (see  \fref{Figtheory}d, top row).
The apical oxygen drains one electron from each of the two vanadium sites in the top layer, whereas the subsurface layers are again close to a V$^{4+}$ configuration except for the  V1 site of the second layer which has $n_{V_{1}}=0.8$.
With the apical oxygen draining all electrons 
from the surface layer, it becomes a dead (insulating) layer. The other layers then essentially behave as the film without apical oxygen but with one layer --the dead one-- less.  Because of the confinement to five layers (six without apical oxygen), quantum well states as shown in \fref{Figtheory}e develop \cite{yoshimatsu_metallic_2011}.

\section{Effect of the surface overoxidation  -- experiment}
\begin{figure*}[hbt]
\includegraphics[width = \linewidth]{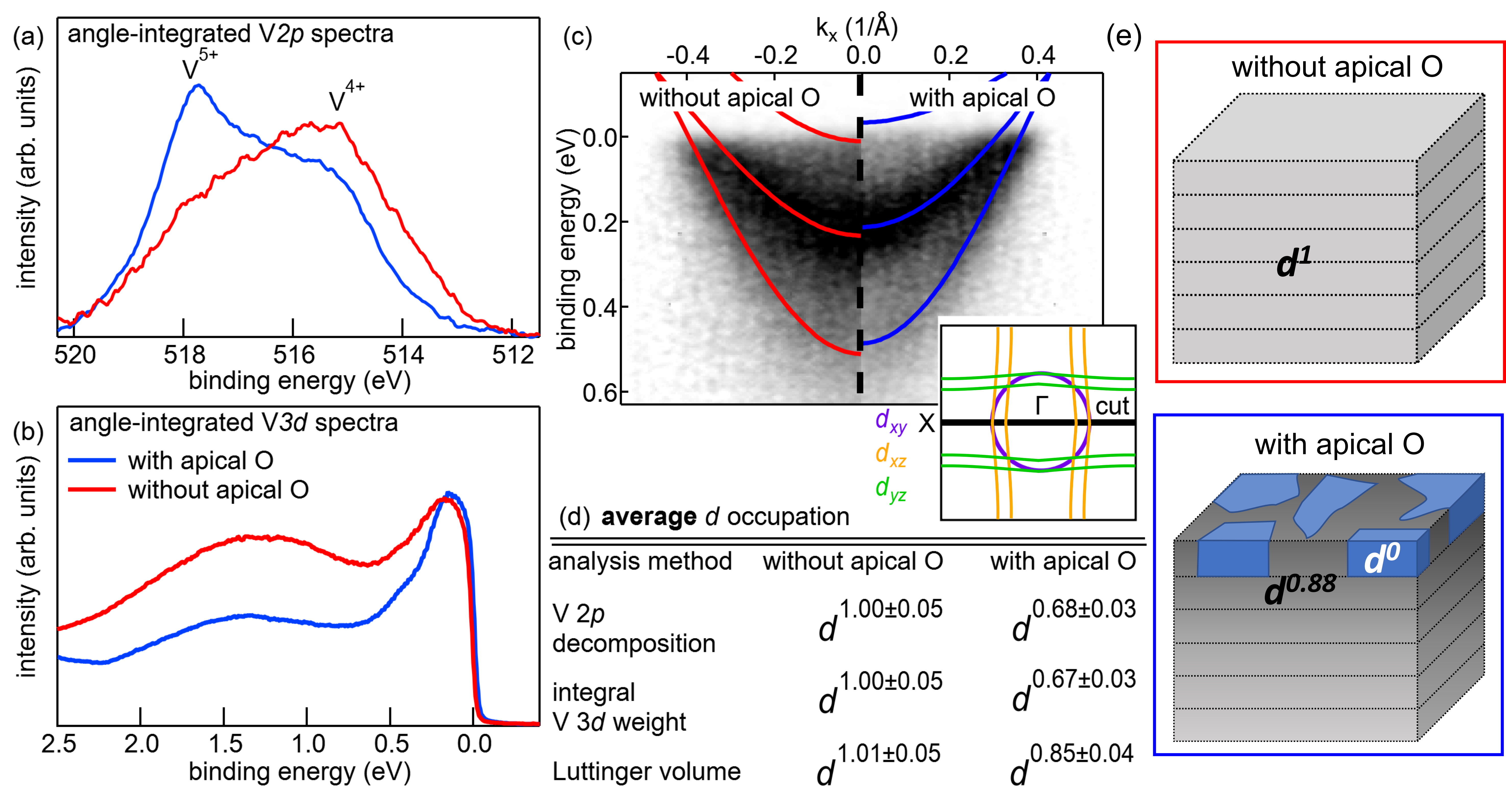}
\caption{
\label{FigARPES}
 Effect of the surface overoxidation on the electronic structure of a 6\,uc SVO film. 
The angle-integrated V\,2$p$ a) and 3$d$ b) spectra, normalized to integration time, change notably between the states with and without apical oxygen coverage which signals a pronounced shift in the average V valence/V\,3$d$ occupation.
c) The angle-resolved V\,3$d$ spectra for a film without (left) and with (right) apical oxygen hardly differ indicating a barely changing band filling in the metallic domains. 
d) The $d$ occupations estimated by different  methods indicate that the $d$ occupation is homogeneous for the clean surface while domains with different $d$ occupation form when apical oxygen adsorbs  as sketched in e).  
 }
\end{figure*}

To investigate experimentally the effect of the apical oxygen on the electronic structure,  a 6\,uc SVO/STO film was probed by soft x-ray photoelectron  spectroscopy  with the degree of surface overoxidation being varied  in a controlled manner.
These measurements were performed with synchrotron radiation for two reasons. First, due to the tunability of the synchrotron light, core levels and valence band can be probed at the same photoelectron kinetic energy and thus the same probing depth. 
We chose low photoelectron kinetic energies around 105\,eV, at which the inelastic mean free path of the photoelectrons amounts to only 3.5\AA~, ensuring a high sensitivity to the topmost SVO layers (see Section 3 in the Supporting Information for the estimate  of the inelastic mean free path).
Second, with the intense hard x-ray beam available at a third generation synchrotron the apical oxygen can be effectively removed through photon-induced desorption \cite{knotekIonDesorptionCorehole1978}, thereby restoring a clean surface (see Methods). 
Using these tactics, we have investigated the sample in two states:
(i) the state immediately after the PLD growth,
which is characterized by additional apical oxygen adsorbing on the VO$_2$-terminated surface during the growth in an oxygen background pressure and which remains intact through an ultrahigh vacuum transfer to the synchrotron, 
and (ii) the state after removal of the apical oxygen by irradiation of hard x-rays.
The corresponding V\,2$p$ spectra of the two states are presented in Figure \ref{FigARPES}a, where the presence and removal of apical oxygen for the states before (blue) and after (red) the hard x-ray exposure is confirmed by the pronounced and nearly quenched V$^{5+}$ components, respectively, at 518\,eV binding energy. \\
To quantify the effect of the apical oxygen on the electronic structure, we determine the V\,3$d$ occupations in both states using three different methods. 
First, from a decomposition of the V\,2$p_{3/2}$ spectral weight into the V$^{4+}$ and V$^{5+}$ contributions we obtain and indicate in Figure  \ref{FigARPES}d what we call  the average  $d$ occupation in the probed volume, \emph{i.e.}, the equivalent $d$ occupation of a homogeneous SVO film. 
Based on the absence of a V$^{5+}$ component in the  V\,2$p_{3/2}$ spectra, the $d$ occupation of the film without apical oxygen is readily  identified as   $d^1$. 
When  the surface is decorated with apical oxygen we find that the average $d$ occupation is lowered to  $d^{0.68}$ (see Section 4  and Figure 5a in the Supporting Information for details).    \\
An independent estimate of the average $d$ occupation can be obtained from the total (angle-integrated) V\,3$d$ weights for the two states, normalized to integration time.
The corresponding spectra are depicted in Figure \ref{FigARPES}b. The V\,3$d$ valence states span the first 2\,eV below the Fermi level  and display the characteristic line shape of a strongly correlated metal, featuring an incoherent lower Hubbard band (LHB)  at about 1.5\,eV and a metallic quasiparticle (QP) state at the Fermi level \cite{fujimori_evolution_1992}.
We find that the overall V\,3$d$ weight drops by 33\% when the clean surface is compared to that with apical oxygen, in line with the 32\%  decrease in the average $d$ occupation inferred from the V\,2$p$ spectra (see Section 4 and Figure 5b in the Supporting Information for details).
Note that the presence or absence of apical oxygen not only  affects the total intensity but also   the line shape of the angle-integrated V\,3$d$ spectrum. \\ 
Yet another way of probing the V\,3$d$ band filling utilizes the determination of the Luttinger $k$-space volume enclosed by the Fermi surface. 
For this purpose, we evaluate the momentum-dependent band structure recorded by angle-resolved photoemission spectroscopy (ARPES). 
Figure \ref{FigARPES}c shows band maps measured along the Brillouin zone cut sketched in the inset for the states with and without apical oxygen. 
Assuming a 6\,uc homogeneous SVO film, we derive the average $d$ band fillings from the observed Fermi wavevectors and the known symmetry and shape of the Fermi surface  to be $d^{1.01}$ and $d^{0.85}$ for the states without and with apical oxygen, respectively, the values of which are also included in the table in Figure  \ref{FigARPES}d for comparison (see Section 4 and Figure 5c-5e in the Supporting Information for details). \\
For the sample state of a clean surface, the table in Figure \ref{FigARPES}d shows that all three methods yield the same result, namely, essentially a $d^1$ occupation as expected from the nominal electron configuration of vanadium in stoichiometric SrVO$_3$ (see top sketch in Figure \ref{FigARPES}e). In contrast, the resulting $d$ occupations diﬀer noticeably for the state with apical oxygen: 
while we  observe larger than 30\%  reductions in the $d$ occupations inferred from the angle-integrated V\,3$d$ and V\,2$p$ spectra, the Luttinger volume derived from the angle-resolved V\,3$d$ measurements decreases by a much smaller amount of only 16\% upon the addition of surface apical oxygen.
In general, such a discrepancy is indicative of an electronic phase separation into metallic and insulating regions on mesoscopic or microscopic length scales, as also observed for SrTiO$_3$ \cite{dudy_situ_2016}.
This is because the band filling extracted from ARPES mapping  is only susceptible to the metallic domains (as insulating regions do not contribute to the Fermi surface), whereas the angle-integrated V\,3$d$ and V\,2$p$ spectra average the band filling over both insulating \textit{and} metallic domains and are insensitive to the spatial distribution of the $d$ electrons. \\

In light of this phase separation scenario, we consider the coexistence of insulating $d^0$ and metallic $d^{1-x}$  domains for the sample state with apical oxygen in line with the experimental results summarized in Figure \ref{FigXPSthickfilm} and the theoretical findings for the layer-resolved orbital occupations (see Figure \ref{Figtheory}d).
We model the SVO film by an inhomogeneous phase separation where laterally scattered $d^0$ domains are confined to the topmost layer and embedded in a homogeneous metallic matrix of band filling $d^{1-x}$ that extends 6\,uc in depth, as shown in the bottom sketch of Figure \ref{FigARPES}e.
We vary $x$ and the surface coverage of the $d^0$ domains to yield the V$^{5+}$/V$^{4+}$ ratio and Luttinger volume that are consistent with the corresponding measured V\,2$p$ spectrum and ARPES band map. 
Note that, while modelling the V\,2$p$ core level intensity, we take into  account the exponential damping of the photoelectrons from sub-surface layers. We reproduce the experimental results for the 6\,uc SVO film in the state with apical oxygen with a $d^{0.88\pm0.05}$ band filling for its metallic regions and (35$\pm7$)\% of its surface  turned into a $d^0$ dead layer.
It should be noted that the probing depth of our photoemission experiments amounts to about 10.5 \AA~(3-times the photoelectron inelastic mean free path given above), and sets a limit on the depth sensitivity of the angle-integrated measurements, and hence the above  estimates are most relevant to  the three topmost SVO layers. \\

Our findings of apical oxygen altering the electronic structure are at variance with the interpretation put forward by Backes \etal\cite{backes_hubbard_2016} of their synchrotron-based ARPES results on SVO films. 
They observe strong changes in the spectral line shape as a function of UV irradiation  and attribute them to beam-induced oxygen \emph{vacancies} forming in  the SVO lattice. 
This interpretation is incompatible with our data, where the x-ray induced nearly 50\% increase of V\,3$d$ spectral weight (Figure \ref{FigARPES}b) is accompanied by a massive V\,2$p$ spectral weight transfer from the V$^{5+}$ to the V$^{4+}$ component (Figure \ref{FigARPES}a), rather than any noticeable further reduction of V (from V$^{4+}$ to V$^{3+}$) expected for oxygen vacancy formation. 
Note that their samples showing the large changes in spectral weight were transferred in air and then annealed in vacuum to recover a clean surface. 
We suspect that the annealing process did not remove all the excess surface oxygen,  which then desorbed under the intense UV light, as reported in this work. 
Despite the different interpretations we observe essentially the same dependence of the V\,3$d$ spectral function on radiation dose.\\ 

\section{Discussion} 
\begin{figure*}[hbpt]
\includegraphics[width = 0.73\linewidth]{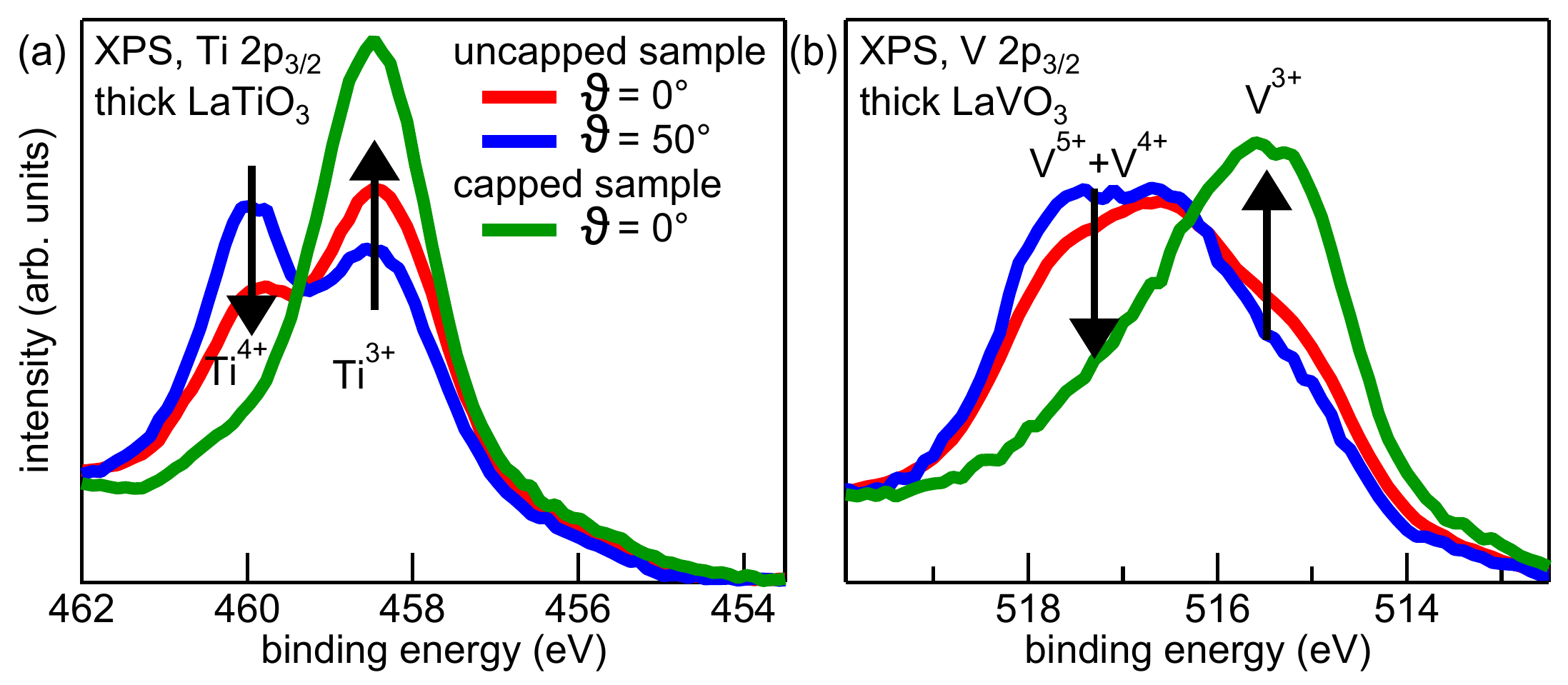}
\centering
\caption{
\label{FigLTOLVO}
a) Ti\,2$p$  spectra of   $\approx$40\,uc thick LaTiO$_3$  films and b) V\,2$p$ spectra of  $\approx$10\,uc thick LaVO$_3$ films.     
The pronounced Ti$^{4+}$, V$^{4+}$ and V$^{5+}$ signal components in the uncapped  films reveal the presence of apical oxygen at the surface. The surface overoxidation is prevented by adding an epitaxial LaAlO$_3$ capping layer.   
}
\end{figure*}

Viewed together, our theoretical and experimental results provide a detailed picture of the adsorption of oxygen at the apical sites of SVO films and its impact on their electronic properties. 
The DMFT calculations consider the limiting case where every second vanadium in the film surface is decorated with an apical oxygen. They show that at this coverage the 3$d$ shell of the V atoms in the topmost layer is completely depleted while the $d^1$ occupation of all layers beneath remains virtually unaffected. 
For the 75\,uc thick SVO film, whose apical oxygen coverage  matches the supposition of the DMFT calculations, the formation of a complete $d^0$ layer at the surface is indeed observed  experimentally in our depth-dependent XPS measurements.  \\
We also find a $d^0$ phase forming in the surface layer of the 6\,uc SVO film when decorated with apical oxygen. However, it does not cover the entire surface. 
We observe additionally a slight electron depopulation of the metallic regions driven by the presence of apical oxygen. 
The latter observation is elusive to our calculations which assume a full coverage of apical oxygen in an ordered $\sqrt{2}\times\sqrt{2}$ R(45$^{\circ}$) arrangement. 
Here it is interesting to note that the fractional order LEED reflections in Figure \ref{FigThicknessDep}c are weaker for the 6\,uc SVO film than for the 75\,uc one, suggesting that, unlike the much thicker film where the coverage of the $\sqrt{2}\times\sqrt{2}$ R(45$^{\circ}$) reconstruction is more complete, the surface of the 6\,uc SVO film should have regions covered by apical oxygen ions that do not follow the same long range order as well as regions covered by less apical oxygen. 
There, the local lattice distortion and thus the crystal field splitting may be reduced and favor light $p$ doping of the film instead of complete depletion of the $d$ band of the surface layer. \\
Beside the change in the $d$ occupation, we also observe that the  V\,3$d$ spectral shape  begins to evolve upon the desorption of apical oxygen. 
With the V\,3$d$ weight exclusively originating from the metallic domains, this indicates a variation in their electronic properties. 
In particular, the apparent increase in the ratio of the LHB and QP weights may signal increased correlations in the absence of apical oxygen, which might be partly explained by the $d$ occupation being driven closer to an integer filling. 
Furthermore, the topmost layer of the SVO film can exhibit enhanced correlations due to reduced coordination numbers
\cite{liebschSurfaceBulkCoulomb2003, sekiyama_mutual_2004} and thus a different QP to LHB ratio from the bulk. 
The presence of $d^0$ domains in the topmost layers of the film with adsorbed oxygen will suppress this surface contribution and alter the V\,3$d$ spectral function. 
Alternatively, defects in the topmost layer may lead to enhanced scattering and hence broadening of the QP feature. 
When $d^0$ domains develop with the adsorption of apical oxygen, surface defects are shielded by the dead layer, 
scattering becomes less important, and the V $3d$ lineshape will feature a sharper QP peak. \\
We have demonstrated that the presence of apical oxygen can have a strong  bearing on the electronic properties of ultrathin SVO films in terms of the layer-dependent crystal-field splitting, $d$ occupation, number of quantum well states, and orbital composition. In addition, we observe electronic phase separation, in  both lateral and vertical directions, into metallic and insulating domains.  
Since the apical oxygen changes  the electronic structure in the same thickness range in which also the  transition from a correlated metal into a Mott insulating phase \cite{yoshimatsu_dimensional_2010}  occurs, 
it is a natural step to link these two phenomena. 
While we can safely assert that the metal-insulator transition is influenced in many ways by surface chemical effects and that a description
within a simple Mott-Hubbard type scenario falls short, a full understanding of the complex interplay between the thickness-dependent electronic transition and 
the adsorption of apical oxygen requires further studies and is beyond the scope of the present work. \\
In view of the fact that TMO thin films are often grown on the TiO$_2$ terminated SrTiO$_3$ surface which results in a BO$_2$ terminated thin film, apical oxygen is a widespread phenomenon for ultrathin TMO films. \cite{ fuchigamiTunableMetallicityCa2009, schutzElectronicStructureEpitaxial2020, scheiderer_tailoring_2018}.
Figure \ref{FigLTOLVO}a exemplarily illustrates that the presented scenario is also observed  in LaTiO$_3$ \cite{scheiderer_tailoring_2018}  and LaVO$_3$ \cite{stubingerHardXrayPhotoemission2021} thin films.
While stoichiometric LaTiO$_3$ and LaVO$_3$ films would be identified by Ti$^{3+}$ and V$^{3+}$ valence states,   signals characteristic of Ti$^{4+}$ and V$^{4+}$ and V$^{5+}$ valence states are seen in bare LaTiO$_3$ and LaVO$_3$ films.  
As for the SVO, these signals decrease with the electron emission angle, which confirms that they are induced by surface overoxidation. 
Oxygen adatoms are likely to also show up at the surface of other non-perovskite crystal structures at which the metal-oxygen ligand octahedron  is broken.\cite{Window2011,Feiten2015,Lantz2015,wahilaBreakdownMottPhysics2020}  
Figure \ref{FigLTOLVO} also demonstrates
an effective way of preventing the surface overoxidation:  
capping the films with an isostructural and inert epitaxial oxide layer which restores the correct stoichiometry and stabilizes the pristine bulk valence states.\\

\section{Conclusion}
Using the example of SVO we demonstrated that apical oxygen, which decorates the surface of a TMO film may severely alter the intrinsic electronic structure in  a multitude of ways. 
The apical oxygen  can lead to the formation of an electronically and magnetically dead layer with a $d^0$ occupation at the surface and can furthermore change the strength of electronic correlations, the scattering and the band filling in the  metallic layers. 
These findings illustrate that the detailed oxygen configuration at the surface, including surface apical oxygen but also potential capping layers,   has to be considered for a  microscopic understanding of transition metal oxide thin films in experiment and theory. 
It is, in particular, essential for the design and operation of nanoscale devices.


\section{Methods}
The SVO thin films were deposited by pulsed laser deposition (PLD) on the TiO$_2$-terminated (001) surfaces of STO substrates, which were prepared following reference \cite{koster_quasi-ideal_1998}. Every substrate was checked by atomic force microscopy  before growth and exhibited an atomically flat surface with terrace steps of single unit cell height.
The SVO films were fabricated  in a high vacuum (oxygen background pressure $<1\cdot10^{-7}$\,mbar)  at a substrate temperature of 650\,$^\circ$C by ablation from a  polycrystalline SrVO$_{3+x}$ pellet employing a KrF excimer laser at a laser fluence of 0.50\,J/cm$^2$ and a pulse repetition rate of 10\,Hz. The target-substrate distance  amounted to 54\,mm. 
The growth of the SVO films was monitored by reflection high-energy electron diffraction. The high sample quality was confirmed by x-ray diffraction and electric transport measurements (see Section 1  in the Supporting Information). \\
The XPS data shown in Figure \ref{FigXPSthickfilm}  and \ref{FigThicknessDep}  was recorded with  a monochromatized Al\,$K_{\alpha}$ laboratory source and an Omicron EA-125 analyzer.
The photoemission  experiments presented in Figure \ref{FigARPES} were conducted at the
two-color beamline I09 at Diamond Light Source. 
The samples were oriented with LEED before the photoemission measurements at the synchrotron.
The V\,3$d$ and V\,2$p$ spectra were probed with linearly polarized soft x-rays  at a photon energy of 105 and 610\,eV, respectively. 
An intense hard x-ray beam at 3\,keV photon energy focused onto the same spot  as the soft x-ray beam\cite{gabel_disentangling_2017} was used to desorb the apical oxygen. The spot size of both x-ray beams is $\approx$ 30$\mu$m $\times$ 50$\mu$m.
The spectra were acquired with a EW4000 photoelectron analyzer (VG Scienta, Uppsala, Sweden) equipped with a wide angle acceptance lens (acceptance angle parallel to analyzer slit 60$^{\circ}$, perpendicular to analyzer slit 0.3$^{\circ}$).
An inelastic mean free path of 3.5\AA~and 18\AA~was used to model the V\,2$p$ spectra measured at a photoelectron kinetic energy of 105\,eV and 970\,eV, respectively.

In DFT the heterostructures were prepared with $n$ layers SrVO$_3$ on top of 5 layers SrTiO$_3$ oriented in the (001)-direction where the final termination of the surface SrVO$_3$ layer is VO$_2$.
The structure without apical oxygen was set up with the constrained in-plane lattice constant of bulk SrTiO$_3$ ($a=b=a_{SrTiO_3}$) while the structure
with apical oxygen was implemented as a $\sqrt{2}\times\sqrt{2}$ supercell ($a=b=\sqrt{2}\;a_{SrTiO_3}$) with every other surface vanadium being connected to an additional apical oxygen atom. 
Above the film, a sufficiently large vacuum layer of roughly $20$\AA\ was added in (001)-direction.
The atomic positions of  the SrTiO$_3$ layer furthest away from the surface SrVO$_3$ were kept fixed, all other atomic positions were completely relaxed using \verb=VASP=\cite{PhysRevLett.77.3865,PhysRevB.47.558,PhysRevB.49.14251,KRESSE199615}.
The full-electron DFT calculation of these structures were then performed with \verb=WIEN2k=\cite{wien2k}, where we verified the consistency of the internal forces between the two codes.
All DFT calculations employed the PBE exchange-correlation potential\cite{PhysRevLett.77.3865}.
The down-folding to a localized Hamiltonian was done with the \verb=Wien2Wannier=\cite{wien2wannier} interface to \verb=Wannier90= \cite{wannier90}. Here we projected purely
onto the $t_{2g}$ orbitals of the $n$ ($2n$) inequivalent vanadium sites of the structures without (with) apical oxygen.
The resulting Hamiltonian was then supplemented with a layer-independent SU$(2)$-symmetric Kanamori interaction $U=U_{intra}=5.25$eV, $J=0.75$eV, $V=U_{inter}=U-2J=3.75$eV.
These values were chosen to deviate slightly from the more typical bulk SrVO$_3$ values in literature\cite{PhysRevLett.92.176403} ($U=5$eV, $J=0.75$eV, $V=3.5$eV) to mimic slightly smaller screening effects in the quasi-two-dimensional structure\cite{zhong_electronics_2015}.
We performed the DMFT calculations at room temperature ($\beta=40$eV$^{-1}$; $T=290$K) with a continuous-time quantum Monte Carlo solver, employing a orbital-diagonal hybridization function, using \verb=w2dynamics=\cite{w2dynamics}.
To obtain the spectral functions on the real frequency axis, the resulting Green's functions on the Matsubara axis were analytically continued using the maximum entropy method\cite{maxent, PhysRevLett.122.127601} implemented in \verb=ana_cont=\cite{ana_cont}.

\medskip
\textbf{Acknowledgements} \par 
The authors gratefully acknowledge funding support from the Deutsche Forschungsgemeinschaft (DFG) through the Würzburg-Dresden Cluster of Excellence on Complexity and Topology in Quantum Matter “ct.qmat” (EXC 2147, Project ID 390858490) as well as through the Collaborative Research Center SFB 1170 “ToCoTronics” (Project ID 258499086) and from  the Austrian Science Fund (FWF) through grants P 30819, P 30997, P 32044,   and P 30213.
 We also wish to thank Diamond Light Source for time on beamline I09 under proposals SI23737 and   SI25151 and  D. McCue for his superb technical support at the I09 beamline.
The authors gratefully acknowledge access to the Vienna Scientific Cluster (VSC) and thank the Gauss Centre for Supercomputing e.V.  (www.gauss-centre.eu) for providing computing time on SuperMUC-NG at the Leibniz Supercomputing Centre (www.lrz.de).\\
J.G. and M.P. contributed equally to this work.

\medskip
\textbf{Supporting Information} \par 
Supporting Information is available from the Wiley Online Library or from the author.

\bibliographystyle{MSP}
\bibliography{SVO_overoxidation_Lit,additional_refs}

\end{document}